\def\deltabarf{\delta\hspace{-.5mm}\raisebox{4.6mm}{{\special{em:graph opens.pcx}}\,\,\,}}

\def\deltabarf{{\sla{\hspace{.5pt}\delta}}}
\def\deltabarf{{\bar\delta}}
\def\deltabarf{{\delta^S}}

\def\deltabar{~{\raisebox{.35em}{-}\hspace{-.44em}\delta }}

\def\aut#1{#1}
\def\ins#1{}

\def\s{ \sigma}

\def\comment#1{}

\def\cm#1{}

\def\>{\rangle}
\def\<{\langle}
\def\comment#1{}

\documentstyle[11pt,appb,epsf]{article}

\title{\Large \bf Nonholonomic Mapping Principle
for Classical Mechanics
in Spactimes with Curvature and Torsion. New Covariant Conservation Law\\
for Energy-Momentum Tensor%
\thanks{
kleinert@physik.fu-berlin.de,
~ http://www.physik.fu-berlin.de/\~{}kleinert }}

\author{
Hagen KLEINERT \\
Institute for Theoretical Physics, FU-Berlin,
Arnimallee 14, D-14195, Berlin, Germany
}

\begin{document}
\maketitle
\begin{abstract}
The lecture explains the geometric basis for
the recently-discovered
nonholonomic mapping
principle which specifies
certain laws of nature in spacetimes with curvature and torsion
from those in flat spacetime, thus replacing and
extending Einstein's equivalence principle.
An important consequence
is a new
action
principle
for
determining the equation of motion of a free spinless point particle
in such spacetimes. Surprisingly,
this equation
contains a torsion force,
although the
action involves only the metric.
This force
changes geodesic into
autoparallel
trajectories, which
are
a direct manifestation of inertia.
The geometric
origin of the torsion force is
a closure failure of parallelograms.
The torsion force changes the covariant conservation
law of the energy-momentum tensor whose
new form is derived.
\end{abstract}

\section{Introduction}

According to Einstein's equivalence principle, gravitational forces
in a small region of spacetime labeled by coordinates $q^\mu$ ($\mu=0,1,2,3$)
can
be removed
by going into locally accelerated coordinates $x^a $ ($a=0,1,2,3$)
by means of a general coordinate transformation $x^a=x^a(q)$,
such as a freely falling elevator.
A mass point lying at
the center of mass of the elevator
does not feel any force.
It undergoes no acceleration
satisfying the Newton-Einstein
equation of motion
$\ddot x^a=0$ ($a=0,1,2,3$), where the dot denotes the derivative with respect to
the proper time $ \sigma $.
This observation has enabled Einstein
to find the physical laws in
curved spacetimes knowing those in flat spacetime.
He simply transformed equations of motion
from the locally Minkowskian
coordinates $x^a$
to the original curvilinear coordinates $q^\mu $,
and postulated the resulting equations
to describe correctly the motion
in a spacetime with gravitational forces.

In this procedure, the elimination of forces
 is not perfect: it holds only
at a single point, the center of mass of the elevator,
not in its neighborhood, which is affected by
tidal forces
that cannot be removed in this way.

Mathematically, however, it is possible
to remove the tidal forces
 with the help of a coordinate transformations $x^a=x^a(q)$
whose derivatives
$ \partial _\mu x^a(q)$ do not satisfy the Schwarz integrability criterion,
i.e., they do not possess commuting derivatives.
Such
transformations cannot be performed in the laboratory, since
the corresponding falling elevator
would not exist.
The spacetime formed by the image coordinates $x^a$
would have defects, as we shall see below,
but it could be chosen to be free of tidal forces
in the neighborhood of the center of mass.
If we admit such mathematical transformations,
Einstein's equivalence principle
can be formulated as a mapping principle:
The correct equations of motion in the presence of gravitational
forces can be obtained by finding in an entire small
 neighborhood of a point $q^\mu$ a local set of coordinates
$x^a$ with the above non-Schwarzian property
and a complete absence of forces,
and by simply mapping the
force-free trajectories
 in these coordinates back into the original spacetime $q^\mu$.

This mathematical procedure opens up the possibility
for discovering the form of physical laws in more general spacetimes.
For instance, we may assume the existence of gravitational forces,
which can only be removed by transformations
$x ^a=x^a(q)$ whose derivatives do not commute.
Such forces are outside of
Einstein's theory.
In fact, such transformations
can be used to remove forces in an equation of motion,
which appear in an
Einstein-Cartan
spacetime with torsion.
By postulating that the images of the force-free trajectories in $x^a$-spacetime
are the correct trajectories in $q^\mu$-spacetime we obtain
the equations for the straightest lines or autoparrallels
in $q$-spacetime, thus contradicting present theories
which find shortest lines or geodesics
for the particle trajectories.

Physically, autoparallel trajectories
may be interpreted as a manifestation of inertia,
which makes particles run along the straightest lines
rather than the shortest ones as generally believed
\cite{Utiyama,Kibble,Hehl1,Hehl2,Hehl3}.
In the absence of torsion, the
 two lines happen to be the same, but in the presence
of torsion
it is hard to conceive, how a particle should know where to go
to make the trajectory to a distant point the shortest curve.
This seems to contradict our concepts of locality.

Nonholonomic mappings
have been of great use in the physics of vortices and defects
in superfluids and crystals \cite{Kleinert1I,Kleinert1,Camb,Bilby,Kroener1,Kroener2,bausch}.
They are also essential for solving
the path integral of the hydrogen atom \cite{Kleinert4}
via the Kustaanheimo-Stiefel transformation \cite{KS}.
In my lecture, I have first shown  how multivalued gauge transformations
can be used to generate magnetic fields
and their minimal coupling to charged particles \cite{Monopol1,Monopol2,Monopol3}.
This part is omitted in these printed notes
to comply with page limitations \cite{details}.
These
multivalued gauge transformations
carry over to
geometry by introducing multivalued
infinitesimal local coordinate transformations
producing infinitesimal curvature and torsion
in a flat background spacetime.

\section{Nonholonomic Mapping Principle}

Let
$dq^\mu$ be a small increment of coordinates
in the physical spacetime. This is mapped into a coordinate increment
 $dx^a$
via a
transformation
\cite{Kleinert4,Kleinert2,Kleinert3,Kleinert7}
\begin{equation}
d x^{a} \, = \, e^{\,a}_{\,\,\,\lambda} ( q ) \, d q^{\lambda} \, ,
\label{LO}
\end{equation}
whose matrix elements $e^{\,a}_{\,\,\,\lambda} ( q )$ are {\em multivalued tetrads\/}.
The transformation can be chosen such that  the length
of $dx^a$ is measured by
the Minkowski metric $ \eta_{ab}$, so that
the metric $g_{\mu \nu}(q)$ in the spacetime $q^\mu$
is given by
\begin{equation}
\label{IN}
g_{\lambda\mu} ( q ) =  e^{\,a}_{\,\,\,\lambda} ( q )
e^{\,b}_{\,\,\,\mu} ( q ) \eta_{ab} \, ,
~~~~~e^{\,a}_{\,\,\,\lambda} ( q ) \equiv \partial x^a ( q )
/ \partial q^{\lambda}.
\end{equation}
Parallel transport in $q$-spacetime is performed
with an affine connection
\begin{equation}
\label{connection0}
 \Gamma_{\mu \nu}{}^ \lambda(q)\equiv
 e_{a}^{\,\,\,\lambda} ( q )\partial_{\mu} \, e^{\,a}_{\,\,\, \nu} ( q ) \, =-
  e^{\,a}_{\,\,\, \nu} ( q ) \, \partial_{\mu} \,e_{a}^{\,\,\,\lambda} ( q )  ,
\end{equation}
where $ e_{\,a}^{\,\,\,\lambda} ( q )
$ are reciprocal multivalued tetrads.
Its antisysmmetric part is
the torsion  tensor \cite{Schouten}
\begin{equation}
S_{\mu \nu}{}^ \lambda(q) = \frac 12\left[
\Gamma_{\mu \nu}{}^ \lambda(q) -\Gamma_{ \nu\mu}{}^ \lambda (q)\right],
\label{torsion}
\end{equation}
and its covariant curl the curvature tensor
\begin{eqnarray}
R_{\mu \nu \lambda }{}^\kappa ( q ) &=&
e_{\, a}^{\,\,\,  \kappa} ( q )
\left( \partial _\mu\partial _ \nu-
\partial _\nu\partial _ \mu \right)
e^{\, a}_{\,\,\,  \lambda} ( q )
\nonumber \\&=&
\partial _\mu \Gamma_{ \nu \lambda}{}^ \kappa
-\partial _\nu \Gamma_{ \mu \lambda}{}^ \kappa
-
\Gamma_{\mu \lambda}{}^ \sigma \Gamma_{ \sigma  \nu}{}^ \kappa
+\Gamma_{\nu \lambda}{}^ \sigma \Gamma_{ \sigma  \mu}{}^ \kappa.
\label{RC}\end{eqnarray}
 Note that if we were to live in $x$-spacetime, we would
register $S_{\mu \nu}{}^ \lambda(q)$
as an object of anholonomity. In the coordinate system
$q$, however, $S_{\mu \nu}{}^ \lambda(q)$
is observable as a torsion.

Recall the way in which the affine
connection $ \Gamma_{\mu \nu}{}^ \kappa$
serves to define a covariant derivative of vector fields $v_\mu(q),~v^\mu(q)$:
\begin{equation}
 D_\mu v_ \nu(q)=\partial _\mu v_ \nu(q)- \Gamma_{\mu \nu}{}^ \lambda (q)v_ \lambda(q),~~~~~
 D_\mu v^  \lambda(q)=\partial _\mu v^ \lambda(q)+ \Gamma_{\mu \nu}{}^ \lambda(q)
v^ \nu(q).~~~~~
\label{covderxb1}\end{equation}
If we lower the last index of the affine connection
by a contraction, $ \Gamma_{\mu \nu \lambda}\equiv g_{ \lambda \kappa} \Gamma_{\mu \nu}{}^ \kappa$,
there exists the decomposition
 \begin{equation}
\Gamma_{\mu\nu\kappa} = \bar{\Gamma}_{\mu\nu\kappa} +
K_{\mu\nu\kappa}                            ,
\label{gamma}
\end{equation}
where
$ \bar{\Gamma}_{\mu\nu\kappa} $
is the Riemann connection, symmetric in $\mu \nu$,
\begin{eqnarray}
 \bar  \Gamma_{\mu \nu \lambda}\equiv  \left\{ \mu \nu ,\lambda \right\} = \frac{1}{2}
            \left( \partial _\mu  g_{\nu \lambda }
            + \partial _\nu  g_{ \lambda\mu }  - \partial _\lambda
            g_{\mu \nu } \right),
\label{2.9}\end{eqnarray}
and
\begin{equation}
K_{\mu\nu\kappa} =
S_{\mu\nu\kappa} - S_{\nu\kappa\mu} + S_{\kappa\mu\nu}\
\label{K}
\end{equation}
 is an antisymmetric tensor in $ \nu \kappa$,
called the
contortion tensor
\cite{Schouten},
formed from the
torsion tensor
 by lowering the last index $S_{\mu\nu\kappa} =
g_{ \kappa\lambda}S_{\mu \nu}{}^ \lambda$.
With the help of the Riemann connection,
we may define another covariant derivative
\begin{equation}
\bar D_\mu v_ \nu(q)=\partial _\mu v_ \nu(q)- \bar\Gamma_{\mu \nu}{}^ \lambda(q) v_ \lambda(q),~~~~~
 \bar D_\mu v^  \lambda(q)=\partial _\mu v^ \lambda(q)+ \bar\Gamma_{\mu \nu}{}^ \lambda(q)
v^ \nu(q).~~~~~
\label{covderxb}\end{equation}

\section{Application to Particle Trajectories}
As an illustration for nonholonomic mappings
to spacetimes with curvature and torsion
we
use the analogy with  defect physics \cite{Kleinert1I,Kleinert1,Camb,Bilby,Kroener1,Kroener2}.%
\begin{figure}[t]
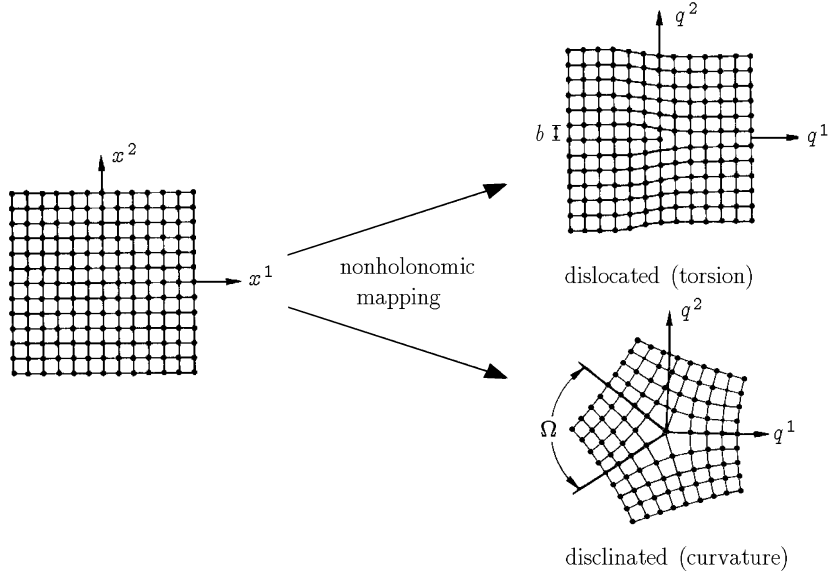

\input dislocc2.tps ~\\[-5mm]
\caption[]{Crystal with dislocation and disclination generated by nonholonomic
coordinate transformations from an ideal crystal.
Geometrically, the former transformation introduces torsion and no curvature,
 the latter
curvature and no torsion.}
\label{torcur}\end{figure}
\begin{figure}[tbh]
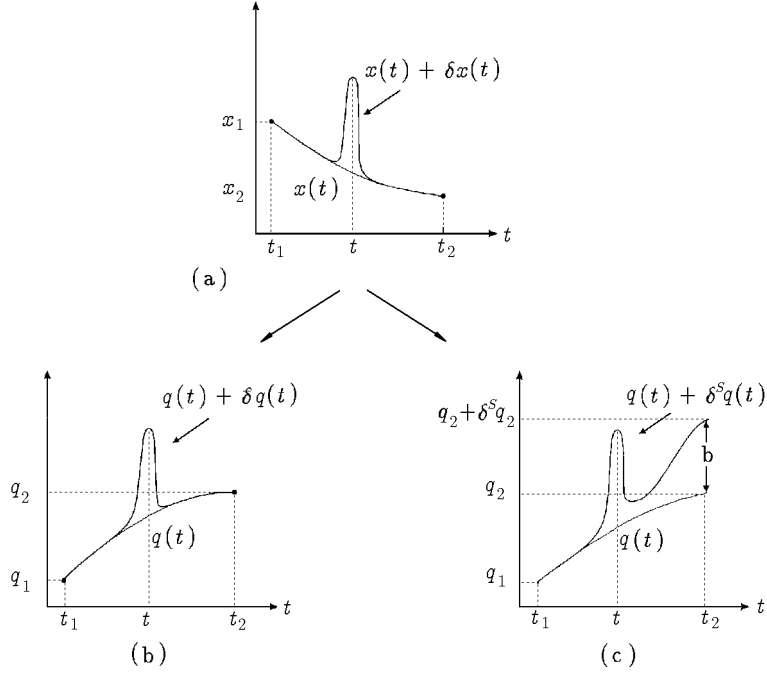

{}\hspace{-2cm}\input nonholm3.tps ~\\
\caption[]{Images
under a holonomic and a nonholonomic mapping
of a fundamental path variation.
In the holonomic case,
the paths $x^a(\s )$ and $x^a(\s )+\delta x^a(\s )$  in (a)
turn into the
paths $q^\mu(\s )$ and $q^\mu(\s ) + \delta q^\mu(\s )$
in (b). In the
nonholonomic case with $S_ {\mu \nu} {}^{\lambda  } \neq 0$,
they go over into
$q^\mu(\s )$ and $q^\mu(\s )+\deltabarf q^\mu(\s )$
shown in (c) with a {closure failure} $ \deltabarf q_2=b^ \mu $ at $t_2$  analogous
to the Burgers vector $b^ \mu$ in a solid with dislocations.}
\label{act}\end{figure}
We consider two types of special
plastic deformations in crystals,
by which
one
produces a single
topological defect
called a disclination (a defect of rotations) and
a dislocation (a defect of translations) (see Fig. \ref{torcur}).
 Just as crystals with dislocations, spacetimes with torsion have a closure failure,
implying that parallelograms do not close. As a consequence,
variations $ \delta q^\mu( \tau )$  of particle trajectories
parametrized arbitrarily by $\tau $
which in the absence of torsion form closed paths,
cannot be zero at both the initial and the final point, but must be
open at the endpoint of the trajectory \cite{Kleinert5,Kleinert6}, as illustrated
 in Fig. \ref{act}.
As a consequence, the Euler-Lagrange equation of spinless point particles
receives a torsion force. This is quite surprising
since the
Lagrangian of a trajectory
$q^\mu(\tau )$,
\begin{equation}
L=-M  \sqrt{g_{\mu \nu}(q(\tau ))\dot q^\mu(\tau )\dot q^\nu(\tau )} ,
\label{rellag@}\end{equation}
(we use natural units with light velocity $c=1$)
contains only the metric \cite{Kleinert6}. The new
Euler-Lagrange equation reads
\begin{eqnarray}
&&
\frac{\partial L}{\partial q^{\,\mu} } -
\frac{d}{d \tau } \frac{\partial L}{\partial
\dot{q}^{\mu} }
= 2  S_{\mu \nu}{}^ \lambda
\dot{q}^{\nu} \frac{\partial
L}{\partial \dot{q}^{ \lambda} }
,\label{EL}
\end{eqnarray}
differing from the standard equation
by
the extra force on the right-hand side
involving the torsion tensor $S_{\mu \nu}{}^ \lambda$.
This extra force
changes geodesic trajectories
into autoparallel ones, whose equation of motion is
\begin{equation}
 \frac{D}{d\s}\dot q^ \nu(\s)\equiv \ddot{q}^\nu(\s)  +
          \Gamma _{\lambda \kappa }{}^\nu(q(\s)) \dot{q}^\lambda(\s)
           \dot{q}^\kappa (\s)=0,
\label{autop}\end{equation}
where
$\sigma$ is the proper time defined by $d\sigma =  \sqrt{g_{\mu \nu}
d q^\mu
d q^\nu
}        $.
The geodesic equation
without the extra force
would contain only the Riemann part of the connection  and reads
\begin{equation}
 \frac{\bar D}{d\sigma}\dot q^ \nu(\sigma)\equiv \ddot{q}^\nu(\sigma)  +
          \bar\Gamma _{\lambda \kappa }{}^\nu(q(\sigma)) \dot{q}^\lambda(\sigma)
           \dot{q}^\kappa (\sigma)=0.
\label{autopb}\end{equation}

A simple variational principle
 to derive the equation of motion (\ref{EL})
is based on the introduction of an {\em auxiliary nonholonomic variation\/}
$  \deltabar q^\mu(\tau )$
of a particle trajectory $q^\mu(\tau )$
which has the novel property
of not commuting with the $\tau $-derivative $d_\tau \equiv d/d\tau $ \cite{Kleinert6}:
\begin{equation}
  \deltabar d_ \tau  q^\mu( \tau)-d_\tau  \deltabar q^\mu(\tau )
= 2S_{ \nu \lambda}{}^\mu
\dot q^ \nu(\tau ) \deltabar q^ \lambda(\tau ).
\label{COMMUTE}\end{equation}
Then (\ref{EL}) is a direct consequence of the new action principle \cite{Kleinert5,Kleinert6}
\begin{equation}
\deltabar {\cal A}=0.
\label{newactpr@}
\end{equation}

An important consistency check for
the correct equations of motion is based on their rederivation
from the covariant conservation law
for the energy-momentum tensor which, in turn,
is a general property of any field theory invariant
under arbitrary (singlevalued) coordinate transformations.

\section{New Covariant Conservation Law for Energy-Momentum Tensor}
To derive this law, we study
the behavior of the relativistic action
\begin{equation}
{\cal A}=-M\int_{\tau _1}^{\tau _2} d\tau  \sqrt{g_{\mu \nu}(q(\tau ))\dot q^\mu(\tau )\dot q^\nu(\tau )}
\label{relact@}\end{equation}
 under the
infinitesimal versions of general coordinate transformations
\begin{eqnarray}
&&dq^\mu\rightarrow dq'{}^\mu=\alpha^\mu{}_ \nu{} dq^\nu  ;~~~~~
\alpha^\mu{}_ \nu{}\equiv \frac{\partial q'{}^\mu}{\partial q{}^\nu},
\label{coortrf@a}\\
&&dq_\mu\rightarrow dq'{}_\mu= \alpha_\mu{}^ \nu{} dq_\nu;~~~~~
\,\alpha_\mu{}^ \nu{}\equiv \frac{\partial q'{}_\mu}{\partial q_\nu}.
\label{coortrf@}\end{eqnarray}
We shall write them
as local translations
\begin{equation}
q^\mu\rightarrow q'{}^\mu(q)
\equiv q^\mu-\xi^\mu(q),
\label{abprel}
\end{equation}
considering from now on only linear terms in the small quantities $\xi^\mu$.

Inserting (\ref{abprel}) into
(\ref{coortrf@a}) and (\ref{coortrf@}), we have
\begin{eqnarray}
  \alpha ^\lambda{}_\nu(q)  \approx \delta^ \lambda{}_\nu
             - \partial _\nu  \xi ^\lambda  (q),~~~~~~~
   \alpha _\mu {}^\nu  (q)\approx \delta _\mu {}^\nu  + \partial _\mu
          \xi ^\nu  (q)                     ,
\label{2.30}\end{eqnarray}
and find from
\begin{eqnarray}
  e_a{}^\mu  (q) =
        \frac{\partial q{}^\mu }{\partial x^a}\rightarrow  e'{}_a{}^\mu  (q') \equiv
        \frac{\partial q'{}^\mu }{\partial x^a} & = &
         \frac{\partial q'{}^\mu }{\partial q^\nu }
         \frac{\partial q^\nu }{\partial x^a}
 \,= \alpha ^\mu {}_\nu (q) e_a{}^\nu (q)          ,
\label{2.25a}\\
  e^a{}_\mu (q)
= \frac{\partial x^a}
         {\partial q{}^\mu }\rightarrow e'{}^a{}_\mu (q') \equiv  \frac{\partial x^a}
         {\partial q'{}^\mu } & = & \frac{\partial q^\nu }
           {\partial q'{}^\mu} \frac{\partial x^a}{\partial q{}^\nu }
  = \alpha _\mu {}^\nu (q) e^a{}_\nu  (q)\nonumber
\label{2.25}\end{eqnarray}
the infinitesimal changes
of the multivalued tetrads
$e_a{}^\mu (q)$ and $e^a{}_\mu (q)$:
\begin{eqnarray}
   && \!\!\!\!\!\!\!
 \delta_E e_a{}^\mu  (q)\equiv
   e'{}_a{}^\mu  (q)
-  e{}_a{}^ \lambda  (q)
    =
 \xi ^\lambda  (q)
\partial _\lambda e_a{}^\mu (q)
                 - \partial _\lambda  \xi ^\mu (q)
 e_a{}^\mu
                 (q),\\  &&\!\!\!\!\!\!\!
 \delta_E e^a{}_\mu (q)\equiv
 e'{}^a{}_\mu (q)-
 e{}^a{}_\mu (q)=
  \xi ^\lambda  (q)
\partial _\lambda
               e^a{}_\mu (q) + \partial _\mu  \xi ^\lambda   (q)
               e^a{}_\lambda (q).
\label{2.31}\end{eqnarray}
The subscript of $\delta_E$ indicates that these changes
are caused by
the infinitesimal versions
of the general coordinate
transformations
introduced by Einstein.

To save parentheses, differential operators are supposed to act only on the
expression after it.
Inserting (\ref{2.31}) into (\ref{IN}),
we obtain the corresponding transformation law for the metric tensor
\begin{equation}
 \delta_E g_{\mu \nu}(q)=
\xi^ \lambda(q)\partial _ \lambda g_{\mu \nu}(q)
+\partial _\mu\xi^ \lambda  (q)g_{ \lambda \nu}(q)
+\partial _\nu\xi^ \lambda  (q)g_{ \mu\lambda }(q)  .
\label{mettr@}\end{equation}
With the help of the
covariant derivative
(\ref{covderxb}),
this
can be rewritten as
\begin{equation}
 \delta_E
g_{\mu \nu}(q)
=\bar D_\mu\xi_ \nu(q)
+\bar D_\nu\xi_ \mu(q).
\label{deltaeg@}\end{equation}
For the
 coordinates $q^\mu$ themselves, the infinitesimal transformation is
\begin{equation}
\delta_E q^\mu=-\xi^\mu(q),
\label{deltaeq@}\end{equation}
which is just the initial transformation (\ref{abprel}) in this notation.

We now calculate the change of the action (\ref{relact@})
under infinitesimal Einstein transformations:
\begin{equation}
 \delta_E
{\cal A}=  \int d^4q
\frac{ \delta{\cal A}}{ \delta g_{\mu \nu}(q)} \delta_E    g_{\mu \nu}(q)
+\int d\tau
\frac{ \delta{\cal A}}{ \delta q^{\mu}(\tau )} \delta_E    q^{\mu }( \tau ).
\label{deltaeA@}\end{equation}
The functional derivative
 $ \delta{\cal A}/ \delta g_{\mu \nu}(q)
$ is the general definition of the energy-momentum tensor
of a system:
\begin{equation}
\frac{ \delta{\cal A}}{ \delta g_{\mu \nu}(q)} \equiv - \frac{1}{2}\sqrt{-g(q)}\, T^{\mu \nu}(q),
\label{@}\end{equation}
where $-g$ is the determinant of $-g_{\mu \nu}$.
 For the spinless particle at hand, the energy-momentum tensor
becomes
\begin{equation}
T^{\mu \nu}(q)=\frac{1}{ \sqrt{-g} } M \int d\s \,\dot
q^\mu (\s )
\dot q^\nu (\s ) \,\delta^{(4)}(q-q(\s )),
\label{explenm@}\end{equation}
where $\s$ is the proper time.
Equation (\ref{explenm@}) and the explicit variations
(\ref{deltaeg@}) and (\ref{deltaeq@}),
bring
(\ref{deltaeA@})
to the form
\begin{equation}
 \delta_E
{\cal A}= -\frac{1}{2} \int d^4q    \sqrt{-g} T^{\mu \nu}(q)
[\bar D_\mu\xi_ \nu(q)
+\bar D_\nu\xi_ \mu(q)]-\int d\tau
\frac{ \delta{\cal A}}{ \delta q^{\mu}(\tau )} \xi^{\mu }( q(\tau) ).
\label{deltaeA3@}\end{equation}
A partial integration of the derivatives yields (neglecting boundary terms at infinity
and using the symmetry of $T^{\mu \nu}$)
\begin{eqnarray}
\delta_E
{\cal A}&=& \int d^4q  \,\left\{
 \partial _ \nu
[ \sqrt{-g} T^{\mu \nu}(q)]
\!+ \!\sqrt{-g}
\bar  \Gamma_{ \nu \lambda}{}^\mu(q)  T^{ \lambda \nu}(q)
\right\}
\xi_ \mu(q)
    \nonumber \\
&-&\int d\tau
\frac{ \delta{\cal A}}
{ \delta q^{\mu}(\tau )} \xi^{\mu }( \tau).
\label{deltaeA4@}
\end{eqnarray}
Because of the  manifest invariance of the action under
general coordinate transformations, the
left-hand side has to vanish
for
arbitrary (infinitesimal) functions
$\xi^\mu(\tau )$.
We therefore obtain
\begin{eqnarray}
 &&
\!\!\!\!\!\!\!\!\!\!\!\!\!
\!\!\!\!\!\!\!\!\!\!\!\!\!
\!\!\!\!\!\!\!\left\{ \partial _ \nu
[ \sqrt{-g} T^{\mu \nu}(q)]
+ \sqrt{-g} \bar \Gamma_{ \nu \lambda}{}^\mu  T^{ \lambda \nu}(q)
\right\}
\xi_\mu(q)
\nonumber \\&&
\!-\!\!\int d\tau
\frac{ \delta{\cal A}}
{ \delta q^{\mu}(\tau )}
 \delta^{(4)}(q\!-\!q(\tau  )) \xi^{\mu }( \tau  )=0.
\label{194@}\end{eqnarray}
To find
the physical content of this equation
we consider first a space
without torsion.
On a particle trajectory, the action is extremal,
so that
the second term vanishes, and
 we obtain the
 covariant conservation law:
\begin{equation}
 \partial _ \nu
[ \sqrt{-g} T^{\mu \nu}(q)]
+ \sqrt{-g}\bar \Gamma_{ \nu \lambda}{}^\mu (q) T^{ \lambda \nu}(q)
=0.
\label{consLAW@}\end{equation}
Inserting
(\ref{explenm@}), this becomes
\begin{eqnarray}
&&\!\!\!\!\!\! M \int ds \,[\dot
q^\mu (\sigma)
\dot q^\nu (\sigma)\partial _\nu \delta^{(4)}(q-q(\sigma))
+
 \bar\Gamma_{ \nu \lambda}{}^\mu (q)\dot q^\nu (\sigma)\dot q^ \lambda(\sigma)
 \,\delta^{(4)}(q-q(\sigma))]=0
. \nonumber \\&&
\label{explenm1}\end{eqnarray}
A partial integration turns this into
\begin{eqnarray}
&& M \int d\sigma\,[
\ddot q^\mu (\sigma)
+
\bar \Gamma_{ \nu \lambda}{}^\mu(q) \dot q^\nu (\sigma)\dot q^ \lambda(\sigma)
]\,\delta^{(4)}(q-q(\sigma))=0
.
\label{explenm2}\end{eqnarray}
Integrating this over a small volume around any
 trajectory point
$q^\mu(\sigma)$, we obtain Eq.~(\ref{autopb})
for a geodesic trajectory.

A similar calculation
was used by Hehl \cite{Hehl3} in his derivation of
 particle trajectories
in the presence of torsion.
Since torsion does not appear in the
action, he found that the trajectories to be geodesic.

The conservation  law (\ref{consLAW@})
can be written more covariantly as
\begin{equation}
 \bar D_ \nu T ^{\mu \nu}(q)=0.
\label{covLAW1@}\end{equation}
This follow directly from the identity
\begin{equation}
\frac{1}{ \sqrt{-g} }\partial _ \nu \sqrt{-g}   =\frac{1}{2}g^{ \lambda \kappa }\partial _ \nu g_{ \lambda \kappa}
=\bar  \Gamma_{ \nu \lambda}{}^  \lambda,
\label{@}\end{equation}
and is a consequence of the
rule of partial integration applied to
(\ref{deltaeA3@}), according to which
a
covariant derivative  can be treated in a volume integral
$\int d^4q \sqrt{-g}f(q)\bar D g(q)$ just like an ordinary derivative in
an
euclidean integral $\int d^4 x f(x) \partial _a g(x)$
After a partial integration neglecting surface terms, Eq.~(\ref{deltaeA3@})
goes over into
\begin{equation}
 \delta_E
{\cal A}\!=\!\frac{1}{2}  \int\! d^4q    \sqrt{-g}\left[
 \bar D_ \nu T^{\mu \nu}(q) \xi_ \mu(q)+
\bar D_ \mu T^{\mu \nu}(q) \xi_ \nu(q)\right]
\!-\!\int \!d\tau
\frac{ \delta{\cal A}}{ \delta q^{\mu}(\tau )} \xi^{\mu }( q(\tau) ).
\label{deltaeA3@p}\end{equation}
whose vanishing for all $\xi^\mu(q)$
yields
directly
(\ref{covLAW1@}), if the action is extremal under ordinary variations
of the orbit.

Our theory does not lead to this
conservation law.
In the presence of torsion,
the particle trajectory does not satisfy
${ \delta{\cal A}}/
{ \delta q^{\mu}(\tau  )} =0$, but according to
(\ref{EL}):
\begin{equation}
\frac{ \delta{\cal A}}
{ \delta q^{\mu}(\tau )}
=
\frac{\partial L}{\partial q^{\,\mu} } -
\frac{d}{d\tau } \frac{\partial L}{\partial
\dot{q}^{\mu} }
= 2  S_{\mu \nu}^{\,\,\,\,\,\, \lambda}
\dot{q}^{\nu} \frac{\partial
L}{\partial \dot{q}^{ \lambda} },
\label{198@}\end{equation}
the right-hand side being equal to
$-M \, 2  S_{\mu \nu \lambda}
\dot q^\nu( \sigma )
\dot q^ \lambda( \sigma )       $ if we choose $\tau $ to be the proper
time $ \sigma$.
Inserting this into (\ref{194@}),
equation
(\ref{explenm2}) receives an extra term and becomes
\begin{eqnarray}
&& M \int d\sigma  \,\left\{
\ddot q^\mu (\sigma)
+
[\bar\Gamma_{ \nu \lambda}{}^\mu (q)
+2S^\mu{}_{ \nu \lambda} (q)
]\dot q^\nu (\sigma)\dot q^ \lambda(\sigma)
\right\} \,\delta^{(4)}(q-q(\sigma))                      =0
,\nonumber \\&&
\label{explenm2n}\end{eqnarray}
thus yielding
the correct autoparallel trajectories
(\ref{autop})
for spinless point particles.

Observe that the extra term in (\ref{deltaeA3@p})
can be expressed
via (\ref{198@})
 in terms of the energy-momentum tensor
(\ref{explenm@})  as
\begin{equation}
 \int d^4q \sqrt{-g}\, 2S^\mu{}_{ \nu \lambda}(q)  T^{ \lambda \nu}(q)
\xi_\mu(q)                        .
\label{@}\end{equation}
We may therefore rewrite the
change of the action (\ref{deltaeA3@})
as
\begin{equation}
 \delta_E
{\cal A}= -\frac{1}{2} \int d^4q    \sqrt{-g}\, T^{\mu \nu}(q)
[\bar D_\mu\xi_ \nu(q)
+\bar D_\nu\xi_ \mu(q)-4S^ \lambda{}_{\mu \nu}\xi_ \lambda(q)].
\label{deltaeA3p@}\end{equation}
The quantity in brackets will be denoted by
$\deltabar_E g_{\mu \nu}(q)$, and is equal to
\begin{equation}
\deltabar_E g_{\mu \nu}(q) = D_\mu\xi_ \nu(q)
+ D_\nu\xi_ \mu(q)                      ,
\label{@}\end{equation}
where $D_\mu$ is the
covariant derivative
(\ref{covderxb1}) involving the full affine connection.
Thus we have
 \begin{equation}
 \delta_E
{\cal A}= - \int d^4q    \sqrt{-g}\, T^{\mu \nu}(q)
 D_\nu\xi_ \mu(q).
\label{deltaeA3pp@}\end{equation}
Integrals over invariant
expressions containing the
covariant derivative $D_\mu$ can be
integrated by  parts
according to a rule very similar to that
for the Riemann covariant derivative
$\bar D_\mu$ (which is derived in Appendix A of \cite{details}). After neglecting
surface terms we find
 \begin{equation}
 \delta_E
{\cal A}=  \int d^4q    \sqrt{-g}\, D_\nu^*T^{\mu \nu}(q)
 \xi_ \mu(q),
\label{deltaeA3pp@}\end{equation}
where
$D_\nu^*=D_\nu+2S_{\nu  \lambda}{}^ \lambda$.
Thus, due to the closure failure in spacetimes with torsion,
the energy-momentum tensor
of a free spinless point particle
satisfies the conservation law
\begin{equation}
D_\nu^*T^{\mu \nu}(q)=0.
\label{covLAWx}\end{equation}
This is to be contrasted
with the conservation law
(\ref{covLAW1@}). The difference between the two laws
can best be seen by rewriting
(\ref{covLAW1@})
as
\begin{equation}
D_\nu^*T^{\mu \nu}(q)+2S_{ \kappa }{}^ {\mu }{}_{\lambda}(q) T^{ \kappa  \lambda}(q)=0.
\label{covLAW2@}\end{equation}
This is the form in which the conservation law
has usually been stated
in the literature \cite{Utiyama,Kibble,Hehl1,Hehl2,Kleinert1}.
When written in the form
(\ref{covLAW1@}) it is obvious that
it is satisfied only by geodesic trajectories.

Note that the variation $\deltabar _Eg_{\mu \nu}(q)$ plays a
similar role
in deriving the new conservation law (\ref{covLAW2@})
as the nonholonomic variation
$\deltabar q^\mu(\tau)$ with the noncommutative property
(\ref{COMMUTE})
does in deriving equations of motion
for point particles.
Indeed, we
may rewrite
the transformation
(\ref{deltaeA@}) formally as
\begin{equation}
 \deltabar_E
{\cal A}=  \int d^4q
\frac{ \delta{\cal A}}{ \delta g_{\mu \nu}(q)} \deltabar_E    g_{\mu \nu}(q)
+\int d\tau
\frac{ \delta{\cal A}}{ \delta q^{\mu}(\tau )} \deltabar_E    q^{\mu }( \tau ).
\label{deltaeAx@}\end{equation}
Now the last term vanishes
according to the new action principle
(\ref{newactpr@}),
from which we derived the
autoparallel trajectory (\ref{autop}).

\section{Consequences for Field Theory of Gravitation with Torsion}
The question arises whether the new conservation law
(\ref{covLAWx}) allows for the construction of an extension of
Einstein's field equation
\begin{equation}
\bar G^{\mu \nu}= \kappa T^{\mu \nu}
\label{EEQ@}\end{equation}
to spacetimes with torsion,
where $ \bar G^{\mu \nu}\equiv \bar R_{\mu \nu}-\frac{1}{2}g_{\mu \nu}\bar R_ \sigma{}^ \sigma$
 is the Einstein tensor formed from the
Ricci tensor $\bar R_{\mu \nu}\equiv \bar R_{ \lambda\mu \nu}{}^ \lambda$
in Riemannian spacetime [$ \bar R_{ \mu \nu \lambda}{}^  \kappa$
being the same covariant curl of $\bar  \Gamma_{\mu \nu}{}^ \lambda$
as $R_{ \mu \nu \lambda}{}^  \kappa$
is of
$  \Gamma_{\mu \nu}{}^ \lambda$ in Eq.~(\ref{RC})].

The standard extension of (\ref{EEQ@}) to spacetimes with torsion
\cite{Utiyama,Kibble,Hehl1,Hehl2,Kleinert1}
 replaces the left-hand side
by the  Einstein-Cartan tensor
 $  G^{\mu \nu}\equiv  R^{\mu \nu}-\frac{1}{2}g^{\mu \nu} R_ \sigma{}^ \sigma$
and becomes
\begin{equation}
 G^{\mu \nu}= \kappa T^{\mu \nu}                .
\label{EEQC}\end{equation}
The Einstein-Cartan tensor
$G^{\mu \nu}$ satisfies a Bianchi identity
\begin{equation}
D_\nu^*G_{ \mu}{}^\nu +2S_{ \lambda\mu}{}^{ \kappa }G_ \kappa{}^ \lambda
-\frac{1}{2}S^{ \lambda}{}_{ \kappa}{}^{; \nu} R_{\mu  \nu \lambda}{}^ \kappa=0,
\label{Bianchid1@}\end{equation}
where $S^{ \lambda}{}_{ \kappa}{}^{; \nu} $
is the Palatini tensor defined by
\begin{equation}
S_{ \lambda  \kappa}{}^{;  \nu}\equiv 2(S_{ \lambda \kappa}{}^  \nu
+ \delta_{ \lambda}{}^ \nu S_{ \kappa  \sigma}{}^ \sigma
- \delta_{  \kappa}{}^  \nu S_{ \lambda  \sigma}{}^ \sigma).
\label{palat@}\end{equation}
It is then
concluded
that the energy-momentum tensor
satisfies
the conservation law
\begin{equation}
D_\nu^*T_{ \mu}{}^\nu +2S_{ \lambda\mu}{}^ \kappa T_ \kappa{}^ \lambda
-\frac{1}{2 \kappa}S^{ \lambda}{}_{ \kappa}{}^{; \nu} R_{\mu  \nu \lambda}{}^ \kappa =0.
\label{cons2l@}\end{equation}
For standard field theories of matter, this is indeed true
if the Palatini tensor satisfies the second Einstein-Cartan field
equation
\begin{equation}
S^{ \lambda \kappa;\nu} = \kappa \Sigma ^{ \lambda \kappa;\nu} ,
\label{@}\end{equation}
where
$ \Sigma^{ \lambda \kappa;\nu} $ is the canonical spin density of the matter fields.
A spinless point particle contributes
only to the first two terms in (\ref{cons2l@}),
in accordance with
(\ref{covLAW2@}).

Which tensor will stand on the left-hand side of the field equation
(\ref{EEQC}) if the energy-momentum tensor satisfies the conservation law
(\ref{covLAWx}) instead of (\ref{covLAW2@})?
At present, we can give an answer only for the case of a pure
gradient torsion  \cite{Peln}
\begin{equation}
S_{\mu \nu}{}^ \lambda                      =
\frac{1}{2}[
\delta_{\mu}{}^ \lambda\partial _ \nu   \sigma
-\delta_{\nu}{}^ \lambda\partial _ \mu   \sigma].
\label{gradtor@}\end{equation}
Then
we
may simply replace (\ref{EEQC}) by
\begin{equation}
e^{\sigma }G^{\mu \nu}= \kappa T^{\mu \nu}.
\label{EEQCn}\end{equation}
 Note that for gradient torsion, $G^{\mu \nu}$ is symmetric
as can be deduced from the fundamental identity (which expresses merely the fact that
the Einstein-Cartan curvature  tensor $R_{\mu \nu \lambda}{}^ \kappa$ is the covariant curl
of the affine connection)
\begin{equation}
{D^*}_ \lambda S_{\mu \nu}{}^{; \lambda}=G_{\mu \nu}-G_{ \nu\mu}.
\label{@}\end{equation}
Indeed, inserting
 (\ref{gradtor@})
into
(\ref{palat@}),
we find the
Palatini tensor
\begin{equation}
S_{ \lambda  \mu}{}^{;\kappa}\equiv -2[ \delta_ \lambda{}^ \kappa
\partial _\mu  \sigma -
\delta_ \mu{}^ \kappa
\partial _ \lambda  \sigma].
\label{palatgr@}\end{equation}
This has a vanishing covariant derivative
\begin{equation}
D^*_ \lambda S_{\mu \nu}{}^{; \lambda}=
-2[D^*_ \mu \partial _ \nu   \sigma- D^*_  \nu \partial _ \mu   \sigma]
=2[
S_{\mu \nu}{}^ \lambda \partial _  \lambda   \sigma
-2S_{\mu \lambda}{}^  \lambda  \partial _  \nu   \sigma
+2S_{\nu \lambda}{}^  \lambda  \partial _  \mu   \sigma]
, \label{@}\end{equation}
since  the terms on the right-hand
side cancel
after using
(\ref{gradtor@}) and $S_{\mu   \lambda}{}^  \lambda\equiv S_\mu
=-\frac{3}{2}\partial _\mu \sigma$.
Now
we insert
(\ref{gradtor@}) into the Bianchi identity
(\ref{Bianchid1@}), with the result
\begin{equation}
\bar D_\nu^*G_{  \lambda}{}^\nu
+\partial _ \lambda \sigma G_ \kappa{}^  \kappa
-\partial _  \nu \sigma G_  \lambda{}^  \nu
+2\partial _\nu \sigma R_{ \lambda}{}^ \nu=
0.
\label{Bianchidgr11@}\end{equation}
Inserting
here $R_{ \lambda \kappa}=G_{ \lambda \kappa}-\frac{1}{2}g_{ \lambda \kappa}G_ \nu{}^ \nu$,
this becomes
\begin{equation}
D_\nu^*G_{  \lambda}{}^\nu
+\partial _  \nu \sigma G_  \lambda{}^  \nu
=
0.
\label{Bianchidgr12@}\end{equation}
Thus we find for the gradient torsion
(\ref{gradtor@}) the Bianchi identity
\begin{equation}
D^*_ \nu (e^{ \sigma}G_{ \lambda}{}^ \nu)=0.
\label{@}\end{equation}
This makes the
 left-hand side
of the new field equation (\ref{EEQCn})
compatible
with the
covariant new conservation law (\ref{covLAWx}), just as in Einstein's theory.

The field equation for the
$ \sigma$-field and thus for
the torsion
is still unknown.

\section{Acknowledgement}

I am grateful to Dr. A. Pelster
for many stimulating discussions
and to Drs. G. Barnich, H. von Borzeskowski,
S.V. Shabanov for useful comments.
My graduate student C. Maulbetsch contributed with many
good questions.

\end{document}